\documentclass[twocolumn,showpacs,aps,prb,preprintnumbers,amsmath,amssymb,floatfix]{revtex4}

\usepackage{graphicx}
\usepackage{dcolumn}
\usepackage{bm}
\usepackage{epsfig}

\begin{document}
\date{\today}

\title{ 
Suppression of surface $p$-wave superconductivity in disordered topological insulators 
}

\author{ G. Tkachov  }

\affiliation{ 
Institute for Theoretical Physics and Astrophysics, W\"urzburg University, Am Hubland, 97074 W\"urzburg, Germany
}

\begin{abstract}
The paper proposes a self-consistent Green function description of the induced surface superconductivity 
in a disordered three-dimensional topological insulator (TI) coupled to an $s$-wave superconductor. 
We recover earlier results regarding the induced spin-triplet $p$-wave pairing, 
showing that a mixture of $p$- and $s$-wave pair correlations appears as a result of broken spin-rotation symmetry on the helical surface of the TI. 
Unlike the $s$-wave pairing, the $p$-wave component is found to be suppressed in dirty TIs 
in which the elastic mean-free path is much smaller than the superconducting coherence length. 
The suppression is due to the generic nonlocality of the spin-triplet correlations, 
which makes them strongly dependent on the mean-free path in a disordered system.  
In dirty TIs the induced superconductivity is predicted to be predominantly $s$-wave like.
In cleaner TIs, however, the $p$-wave component may reach a magnitude comparable with (but not larger than) the $s$-wave pairing.       
\end{abstract}
\maketitle

\section{Introduction}
\label{Intro}

Topological insulators (TIs) are a novel class of materials (see e.g. reviews \onlinecite{Hasan10,Qi11}) in which the character of electron transport varies 
from insulating in the interior of the material to metallic near its surface. 
Such an atypical conduction character originates from specific electronic states that, for topological reasons, exist only near the surface of the material. 
The defining property of the TI surface states is their spin helicity whereby the spin of the charge carrier follows the direction of its momentum vector. 
The spin helicity is preserved in collisions with impurities and, generally, with any nonmagnetic crystal or sample defects, due to which TIs have been considered as platforms for intriguing applications, ranging from spintronics to topological quantum information processing. 

One of the recent exciting developments in the TI research is the theoretical prediction~\cite{Fu08} 
of unconventional $p$-wave superconductivity and Majorana states in TI/superconductor (S) junctions 
(see also reviews \onlinecite{Hasan10,Qi11,Tanaka12,Alicea12,Beenakker13a,GT13}).
In three-dimensional (3D) TIs the superconductivity can be induced by depositing an $s$-wave superconductor (e.g. Al, W or Nb) 
on the surface of the TI material.~\cite{D_Zhang11,Koren11,Sacepe11,Veldhorst12,Wang12_STI,Williams12,Maier12,Cho12,Oostinga13}
The unconventional superconductivity arises from the helicity of the surface states, 
which breaks spin rotation symmetry, allowing for mixed singlet $s$-wave and triplet $p$-wave pair correlations.
Theoretical aspects of the superconducting proximity effect in the TIs have been considered in Refs. 
\onlinecite{Stanescu10,Linder10,Potter11,Labadidi11,Khaymovich11,Virtanen12,Yokoyama12,Black12},\onlinecite{Maier12},\onlinecite{GT13}.

\begin{figure}[b]
\begin{center}
\includegraphics[width=80mm]{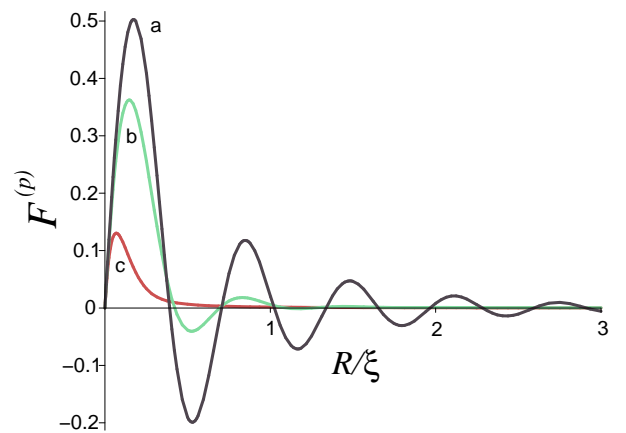}
\end{center}
\caption{
Amplitude of spin-triplet $p$-wave correlations $F^{ (p) }$ versus distance $R$ (in units of coherence length of a clean TI, $\xi$)  
for different disorder strengths: (a) $\xi/\ell = 0$ (clean TI), (b) $\xi/\ell = 5$, and (c) $\xi/\ell = 30$ (dirty TI); 
$\ell$ is elastic mean-free path in TI [see also Eqs.(\ref{F^p}) and (\ref{F^p_1})].
}
\label{F_p_fig}
\end{figure} 

\begin{figure}[t]
\begin{center}
\includegraphics[width=80mm]{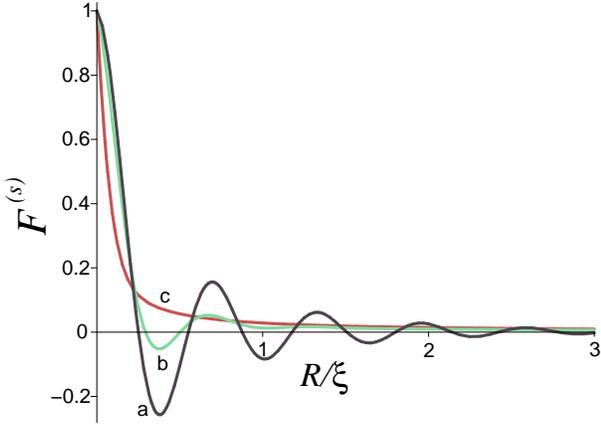}
\end{center}
\caption{
Amplitude of spin-singlet $s$-wave correlations $F^{(s)}$ versus distance $R$ (in units of coherence length of a clean TI, $\xi$)  
for different disorder strengths: (a) $\xi/\ell = 0$ (clean TI), (b) $\xi/\ell = 5$, and (c) $\xi/\ell = 30$ (dirty TI) [see also Eqs.(\ref{F^s}) and (\ref{F^s_1})].
}
\label{F_s_fig}
\end{figure}

This paper addresses the role of elastic impurity scattering in the superconducting proximity effect 
on the surface of a 3D TI. The influence of disorder on superconducting properties of TIs still remains largely unexplored. 
On the one hand, the conservation of the spin helicity in elastic collisions suggests that the spin structure of the pair correlations should remain intact in impure TIs. 
Also, Potter and Lee~\cite{Potter11} have recently proved the robustness of the induced superconducting gap in the TI against elastic impurity scattering. 
A related issue has been discussed in Ref. \onlinecite{Lutchyn12} for a conventional semiconductor with spin-orbit coupling.
On the other hand, the Pauli exclusion makes the spin-triplet pairs inherently nonlocal in space and, hence, dependent on a carrier mean-free path in a disordered system. 
One should therefore expect suppression of the $p$-wave correlations in dirty TIs with the mean-free path smaller than the superconducting coherence length.  
This expectation is confirmed below by direct calculations employing self-consistent Green functions of a disordered S/TI bilayer. 
Concretely, we analyze the real-space amplitudes of the induced p- and $s$-wave pair correlations and 
their evolution from a clean to a dirty TI (see also Figs. \ref{F_p_fig} and \ref{F_s_fig}). 
We also show that these new results are consistent with the earlier conclusions~\cite{Potter11} 
regarding the robustness of the density of states in disordered TIs.

The subsequent sections give a complete account of the theoretical approach adopted in this paper. 
In Sec. II a model for disordered S/TI bilayers is introduced. 
Section III describes a conventional $s$-wave superconductor which is used 
as the source of the proximity effect. 
Section IV is devoted to induced superconductivity in clean and disordered TIs and contains the discussion of the main results.

\begin{figure}[b]
\begin{center}
\includegraphics[width=60mm]{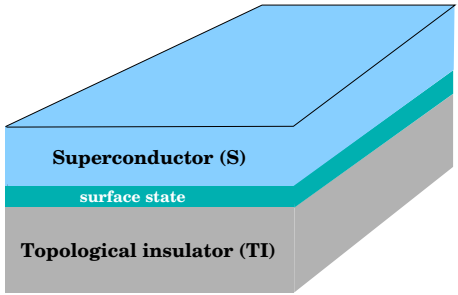}
\end{center}
\caption{
Schematic of a thin-film superconductor (S)/topological insulator (TI) surface contact.
}
\label{Geo}
\end{figure}

\section{ S/TI bilayer model }

\subsection{Hamiltonian}

We begin by reviewing the superconducting proximity effect in a planar interface between a thin singlet $s$-wave superconductor (S) film 
and the TI surface (see also Fig. \ref{Geo}). Such a hybrid system was first considered by Fu and Kane \cite{Fu08}. 
In their approach the proximity effect on the TI surface is described by a phenomenological singlet pairing potential. 
On the other hand, microscopic approaches (e.g. McMillan's model \cite{McMillan68}) 
allow for a more general description of the proximity effect in terms of the Green functions of the S. 
We will employ a microscopic model close in spirit to McMillan's one \cite{McMillan68} and its adaptations to various low-dimensional systems 
(see e.g., Refs. \onlinecite{GT13,Maier12,Stanescu10,Potter11,Golubov04,GT04,GT05,Fagas05,Kopnin11,Stanescu11,Lutchyn12}). 
Without losing essential physics we will treat both the TI surface state and the S film  
as two-dimensional (2D) systems in which electronic states are labeled by the in-plane momentum ${\bm k}$. 
Assuming tunneling coupling between the systems, we can write the Hamiltonian of such a bilayer as follows
\begin{eqnarray}
H=\frac{1}{2}\sum_{ {\bm k},{\bm k^\prime}  }
[ A^\dagger_{\bm k} B^\dagger_{\bm k} ] 
\left[
\begin{array}{cc}
H^{^S}_{ {\bm k} , {\bm k^\prime} } & T^*_{ {\bm k} , {\bm k^\prime} }    \\
T_{ {\bm k} , {\bm k^\prime} } & H^{^N}_{ {\bm k} , {\bm k^\prime} }
\end{array}
\right]
\left[
\begin{array}{c}
A_{\bm k^\prime}
\\
B_{\bm k^\prime}
\end{array}
\right],
\label{H}
\end{eqnarray}
where $A^\dagger_{\bm k}$ and  $B^\dagger_{\bm k}$ ($A_{\bm k}$ and $B_{\bm k}$) are the creation (destruction) operators of the S and TI, respectively, in the Nambu 
(particle-hole) representation:
\begin{eqnarray}
A_{\bm k} = \left[ 
\begin{array}{c} 
 a_{ \uparrow \bm{k} } \\
 a_{ \downarrow \bm{k} } \\
 a^\dagger_{ \uparrow -\bm{k} } \\
 a^\dagger_{ \downarrow -\bm{k} } \\
\end{array}
\right], \qquad
B_{\bm k} = \left[ 
\begin{array}{c} 
 b_{ \uparrow \bm{k} } \\
 b_{ \downarrow \bm{k} } \\
 b^\dagger_{ \uparrow -\bm{k} } \\
 b^\dagger_{ \downarrow -\bm{k} } \\
\end{array}
\right]. 
 \label{A,B}
\end{eqnarray}
The diagonal elements $H^{^S}_{ {\bm k} , {\bm k^\prime} }$ and $H^{^N}_{ {\bm k} , {\bm k^\prime} }$ of the matrix in Eq. (\ref{H}) are 
the Hamiltonians of the superconductor and the normal TI, respectively, in the absence of the tunneling, whereas the off-diagonal matrix element
$T_{ {\bm k} , {\bm k^\prime} }$ describes the tunneling coupling between the two systems. Below we define explicitly these operators.
In the chosen basis (\ref{A,B}) the Hamiltonians $H^{^S}_{ {\bm k} , {\bm k^\prime} }$ and $H^{^N}_{ {\bm k} , {\bm k^\prime} }$ are 
\begin{eqnarray}
H^{^S}_{ {\bm k} , {\bm k^\prime} }=
\left[
\begin{array}{cc}
h^{^S}_{\bm k} \delta_{ {\bm k} , {\bm k^\prime} } + \sigma_{_0} V_{ {\bm k} , {\bm k^\prime} }  & i\sigma_y \Delta_{_S} {\rm e}^{i\chi} \delta_{ {\bm k} , {\bm k^\prime} }    \\
- i\sigma_y\Delta_{_S} {\rm e}^{-i\chi} \delta_{ {\bm k} , {\bm k^\prime} } & -h^{^{S*}}_{ -{\bm k} }\delta_{ {\bm k} , {\bm k^\prime} } - \sigma_{_0} V^*_{ -{\bm k} , -{\bm k^\prime} } 
\end{array}
\right], 
\label{H_S}
\end{eqnarray}
\begin{equation}
h^{^S}_{\bm k} = \left[ A_{_S} {\bm k}^2  - E_{_S} \right] \sigma_{_0},
\label{h_S}
\end{equation}
\begin{eqnarray}
H^{^N}_{ {\bm k} , {\bm k^\prime} }=
\left[
\begin{array}{cc}
h^{^N}_{\bm k} \delta_{ {\bm k} , {\bm k^\prime} } + \sigma_{_0} V_{ {\bm k} , {\bm k^\prime} }  & 0    \\
0 & -h^{^{N*}}_{ -{\bm k} }\delta_{ {\bm k} , {\bm k^\prime} } - \sigma_{_0} V^*_{ -{\bm k} , -{\bm k^\prime} }
\end{array}
\right], 
\label{H_N}
\end{eqnarray}
\begin{eqnarray}
h^{^N}_{\bm k} = A_{_N} \mbox{\boldmath$\sigma$} \cdot {\bm k}  - \sigma_{_0} E_{_N}. 
\label{h_N}
\end{eqnarray}
Here $h^{^S}_{\bm k}$ and $h^{^N}_{\bm k}$ are the normal-state Hamiltonians of the S and N systems 
with corresponding band structure parameters, $A_{_S}$ and $A_{_N}$, and Fermi energies, $E_{_S}$ and $E_{_N}$;~\cite{Fermi}
$\Delta_{_S}$ and $\chi$ are the pairing potential and its phase in the S, $\sigma_{x,y}$ are Pauli spin matrices ($\sigma_{_0}$  is the unit matrix), 
and $V_{ {\bm k} , {\bm k^\prime} }$ is the disorder potential characterized by the correlation function 
\begin{equation}
\langle V_{ {\bm k} , {\bm k^\prime} }V_{ {\bm k}_1 , {\bm k^\prime}_1 } \rangle 
= \frac{u^2}{ a }\, \delta_{ {\bm k} - {\bm k^\prime} , -{\bm k}_1 + {\bm k^\prime}_1}, 
\label{VV}
\end{equation}
where $a$ denotes the contact area.
Finally, the tunneling matrix in the chosen basis is 
\begin{eqnarray}
T_{ {\bm k} , {\bm k^\prime} }=
\left[
\begin{array}{cc}
\sigma_{_0} t_{ {\bm k} , {\bm k^\prime} }  & 0    \\
0 & -\sigma_{_0} t^*_{ -{\bm k} , -{\bm k^\prime} }  
\end{array}
\right]. 
\label{T}
\end{eqnarray}
Since the contact area in such structures is large (of order of $\mu$m$^2$), 
we assume that the tunneling coupling randomly fluctuates on the surface and treat $t_{ {\bm k} , {\bm k^\prime} }$ 
as a random matrix characterized by the correlation function
\begin{equation}
\langle t_{ {\bm k} , {\bm k^\prime} } t_{ {\bm k}_1 , {\bm k^\prime}_1 } \rangle 
= \frac{ \mathcal{T}^2 }{ a } \, \delta_{ {\bm k} - {\bm k^\prime} , -{\bm k}_1 + {\bm k^\prime}_1 }. 
\label{tt}
\end{equation}
Thus, both tunneling $t_{ {\bm k} , {\bm k^\prime} }$ and 
disorder $V_{ {\bm k} , {\bm k^\prime} }$ can be treated simultaneously and on equal footing, 
using the standard self-consistent Born approximation for the Green function of the SN system.

\subsection{Bilayer Green function}

In order to describe the hybrid S/TI system it is convenient to use a matrix Green function:
\begin{eqnarray}
&&
\hat{G}_{ {\bm k} , {\bm k^\prime} }(t,t^\prime) =\frac{1}{i\hbar} 
\left\langle\left\langle  
\left[
\begin{array}{c}
A_{\bm k }(t)
\\
B_{\bm k }(t)
\end{array}
\right]
\otimes  
[ A^\dagger_{\bm k^\prime}(t^\prime) B^\dagger_{\bm k^\prime}(t^\prime) ] 
\right\rangle\right\rangle=
\nonumber\\
&&
=
\left[
\begin{array}{cc}
G^{^S}_{ {\bm k} , {\bm k^\prime} }(t,t^\prime)  &  G^{^{SN}}_{ {\bm k} , {\bm k^\prime} }(t,t^\prime)   \\
G^{^{NS}}_{ {\bm k} , {\bm k^\prime} }(t,t^\prime)  &  G^{^N }_{ {\bm k} , {\bm k^\prime} }(t,t^\prime) 
\end{array}
\right].
\label{G1}
\end{eqnarray}
It involves all direct time-ordered products of $A_{\bm k }(t)$ ($B_{\bm k }(t)$) and  $A^\dagger_{\bm k^\prime}(t^\prime)$ ($B^\dagger_{\bm k^\prime}(t^\prime)$) operators, 
$\langle\langle ...\rangle\rangle$ denotes averaging with the ground-state statistical operator and, simultaneously, averaging over the realizations of random matrices 
$V_{ {\bm k} , {\bm k^\prime} }$ (\ref{VV}) and  $t_{ {\bm k} , {\bm k^\prime} }$ (\ref{tt}). In Eq. (\ref{G1}) the diagonal entries 
$G^{^S}$ and $G^{^N }$ describe the S and TI, respectively, while the off-diagonal ones  $G^{^{SN}}$ and 
$G^{^{NS}}$ are the hybrid Green functions due to the tunneling. All entries are $4\times4$ matrices in basis (\ref{A,B}).

Using the self-consistent Born approximation with respect to both $V_{ {\bm k} , {\bm k^\prime} }$ and  $t_{ {\bm k} , {\bm k^\prime} }$, 
we find the Green function  $\hat{G}_{ {\bm k} , {\bm k^\prime} }(\epsilon) = \hat{G}_{\bm k}(\epsilon)\delta_{ {\bm k} , {\bm k^\prime} }$, 
where $\hat{G}_{\bm k}(\epsilon)$ obeys the Dyson equation
\begin{equation}
\hat{G}_{\bm k}(\epsilon) = \hat{G}^0_{\bm k}(\epsilon) + \hat{G}^0_{\bm k}(\epsilon) \hat{\Sigma}_{\bm k}(\epsilon)\hat{G}_{\bm k}(\epsilon),
\label{Eq_G}
\end{equation}
where $\hat{G}^0_{\bm k}(\epsilon)$ is the Green function of the decoupled disorder-free SN, and  
$\hat{\Sigma}_{\bm k}(\epsilon)$ is the self-energy matrix with the following structure in SN space: 
\begin{widetext}

\begin{eqnarray}
\hat{\Sigma}_{\bm k}(\epsilon) = 
\left[
\begin{array}{cc}
\tau_z \sigma_{_0} [u^2\mathcal{G}^{^S}       + \mathcal{T}^2 \mathcal{G}^{^N}     ]\tau_z\sigma_{_0} & 
\tau_z \sigma_{_0} [u^2\mathcal{G}^{^{SN} }   + \mathcal{T}^2 \mathcal{G}^{^{NS}}  ]\tau_z \sigma_{_0} \\
\tau_z \sigma_{_0} [u^2\mathcal{G}^{^{NS} }   + \mathcal{T}^2 \mathcal{G}^{^{SN}}  ]\tau_z\sigma_{_0} & 
\tau_z \sigma_{_0} [u^2\mathcal{G}^{^N}       + \mathcal{T}^2 \mathcal{G}^{^S}     ]\tau_z\sigma_{_0}
\end{array}
\right], 
\label{Sigma}
\end{eqnarray}
\end{widetext}
In Eq. (\ref{Sigma}) the terms $\propto u^2 $ and $\mathcal{T}^2$ are the disorder and tunneling self-energies, respectively, 
$\tau_z$ is the Pauli matrix in particle-hole space, and $\mathcal{G}^{^{S,N}}$ and $\mathcal{ G }^{^{SN,NS}}$ denote the momentum integrated Green functions: 
\begin{eqnarray}
\mathcal{G}^{^{S,N}} =\int \frac{ d{\bm k} }{(2\pi)^2} G^{^{S,N}}_{\bm k}, \quad 
\mathcal{ G }^{^{SN,NS}} =\int \frac{ d{\bm k} }{(2\pi)^2} G^{^{SN,NS}}_{\bm k}.
\label{G_int}
\end{eqnarray}

\section{ Model of the superconductor }
\label{S}

We now demonstrate that Eqs. (\ref{Eq_G}) and (\ref{Sigma}) recover the known results for impure superconductors~\cite{AGD}
and explicitly calculate the Green function $\mathcal{G}^{^S}$ (\ref{G_int}) which will be needed later for the analysis of the proximity effect. 
Since the tunneling coupling to the TI has only a minor effect on the S, 
we can calculate the S Green function $G^{^S}_{\bm k}$ from the Dyson equation with $\mathcal{T}=0$:   
\begin{equation}
[ \epsilon - H^{^S}_{\bm k} - u^2 \tau_z\sigma_{_0} {\cal G}^{^S} \tau_z\sigma_{_0} ] G^{^S}_{\bm k} ={\bm 1}, \quad 
{\bm 1} = \left[\begin{smallmatrix} \sigma_{_0} & 0\\ 0 &  \sigma_{_0} \end{smallmatrix} \right],
\label{Eq_G_S}
\end{equation}
where
\begin{equation}
H^{^S}_{\bm k}=
\left[
\begin{array}{cc}
h^{^S}_{\bm k} & i\sigma_y \Delta_{_S} {\rm e}^{i\chi}   \\
- i\sigma_y\Delta_{_S} {\rm e}^{-i\chi}  & -h^{^{S*}}_{ -{\bm k} } 
\end{array}
\right]. 
\label{H_S_0}
\end{equation}
For a large Fermi energy $E_{_S} \gg \Delta_{_S}$ and near the S Fermi surface 
the solution of Eq. (\ref{Eq_G_S}) is given by (see e.g. Ref. \onlinecite{AGD})
\begin{eqnarray}
G^{^S}_{\bm k}= 
\frac{
\left[
\begin{array}{cc}
\sigma_0   \, ( \epsilon^\prime_{_S} + \eta_{_S}(k) ) & 
i\sigma_y  \, \Delta^\prime_{_S} \, {\rm e}^{i\chi}  \\
-i\sigma_y \, \Delta^\prime_{_S} \, {\rm e}^{-i\chi} & 
\sigma_0   \, ( \epsilon^\prime_{_S} - \eta_{_S}(k) )
\end{array}
\right] 
}
{
\epsilon^{\prime 2}_{_S} - \eta^2_{_S}(k) - \Delta^{\prime 2}_{_S} 
}, 
\label{G_S}
\end{eqnarray}
where $\eta_{_S}(k)=\hbar v_{_S}(k-k_{_S})$ with $v_{_S}$ and $k_{_S}$ being the Fermi velocity and momentum in the S, 
$\Delta^\prime_{_S }( \epsilon, \hat{\bm k} )$ and  
$\epsilon^\prime_{_S } ( \epsilon, \hat{\bm k})$ are the functions of energy $\epsilon$ and unit vector $\hat{\bm k}$ in momentum direction on the Fermi surface. 
These functions should be calculated self-consistently from the equations:~\cite{AGD}
\begin{eqnarray}
&&
\Delta^\prime_{_S } = \Delta_{_S} + \frac{i\hbar}{ 2\tau_{_S} } 
\overline{
\frac{ \Delta^\prime_{_S} }{  \sqrt{ \epsilon^{\prime 2}_{_S} - \Delta^{\prime 2}_{_S}  }  }
},
\label{Delta_S}\\
&&
\epsilon^\prime_{_S } = \epsilon  + \frac{i\hbar}{ 2\tau_{_S} } 
\overline{
\frac{  \epsilon^\prime_{_S }   }{  \sqrt{ \epsilon^{\prime 2}_{_S }  - \Delta^{\prime 2}_{_S}  }  }
},
\label{E_S}
\end{eqnarray}
where $\overline{( ...)} =\int^{2\pi}_0\frac{  d\phi_{ \hat{\bm k} }  }{2\pi} (...) $ 
is the angle averaging over the momentum direction $\hat{\bm k}$ on the Fermi surface and the time-scale 
\begin{equation}
\tau_{_S}=\hbar/(2 \pi u^2 \nu_{_S} )
\label{tau_S}
\end{equation}
is related to the disorder strength and coincides with the elastic 
life-time. $\nu_{_S}$ is the normal-state density of states (DOS) per spin in the S.

In what follows we will need the momentum-integrated Green function ${\cal G}^{^S}$ (\ref{G_int}) 
which can be obtained from Eqs. (\ref{Eq_G_S}), (\ref{G_S}), (\ref{Delta_S}) and (\ref{E_S}) as  
\begin{eqnarray}
\mathcal{G}^{^S} = -i \pi N_{_S}  
\left[
\begin{array}{cc}
\sigma_0   \, \overline{ g_{_S}( \epsilon, \hat{\bm k} ) } & 
i\sigma_y  \, \overline{ f_{_S}( \epsilon, \hat{\bm k} ) }  \, {\rm e}^{i\chi}  \\
-i\sigma_y \, \overline{ f_{_S}( \epsilon, \hat{\bm k} ) }  \, {\rm e}^{-i\chi} & 
\sigma_0   \, \overline{ g_{_S}( \epsilon, \hat{\bm k} ) }
\end{array}
\right], 
\label{G^S_int}
\end{eqnarray}
where the functions $g_{_S }(\epsilon, \hat{\bm k})$ and $f_{_S }(\epsilon, \hat{\bm k})$ satisfy the equations:
\begin{eqnarray}
&&
g_{_S }(\epsilon, \hat{\bm k} ) 
\left( 
\Delta_{_S} + \frac{i\hbar}{ 2\tau_{_S} } 
\overline{
f_{_S }(\epsilon, \hat{\bm k} ) 
}
\right)
=
\label{Eq1_g_f_S}\\
&&
=f_{_S }(\epsilon, \hat{\bm k} )
\left(
\epsilon  + \frac{i\hbar}{ 2\tau_{_S} } 
\overline{ 
g_{_S }(\epsilon, \hat{\bm k} )
}
\right),
\nonumber\\
&& 
g^2_{_S }(\epsilon, \hat{\bm k} ) - f^2_{_S }(\epsilon, \hat{\bm k} ) = 1. 
\label{norm}
\end{eqnarray}
These equations have isotropic solutions $g_{_S }(\epsilon, \hat{\bm k} ) = g_{_S}(\epsilon)$ and $f_{_S }(\epsilon, \hat{\bm k} )=f_{_S}(\epsilon)$. 
In this case Eq. (\ref{Eq1_g_f_S}) becomes disorder-independent: $g_{_S}(\epsilon) \Delta_{_S} = \epsilon f_{_S}(\epsilon)$. 
Solving it together with Eq. (\ref{norm}) yields the well known result: \cite{AGD}
\begin{eqnarray}
g_{_S}(\epsilon)=\frac{\epsilon}{\sqrt{ \epsilon^2 - \Delta^2_{_S} }},
\qquad 
f_{_S}(\epsilon)=\frac{\Delta_{_S}}{\sqrt{ \epsilon^2 - \Delta^2_{_S} }}.
\label{g_f_S}
\end{eqnarray}
Thus, the momentum-integrated S Green function depends only on energy through functions (\ref{g_f_S}).

\section{ Induced superconductivity in TI }
\label{N}

The induced superconductivity in the TI is described by the Dyson equation for the Green function $G^{^N}_{\bm k}$ [see  Eqs. (\ref{Eq_G}) and (\ref{Sigma})].   
Up to the $\mathcal{T}^2$ order in tunneling, the equation for $G^{^N}_{\bm k}$ is
\begin{eqnarray}
[ \epsilon - H^{^N}_{\bm k} - \mathcal{T}^2 \tau_z\sigma_0 \mathcal{ G }^{^S}  \tau_z\sigma_0 - u^2 \tau_z\sigma_0 \mathcal{ G }^{^N}  \tau_z\sigma_0 ] G^{^N}_{\bm k} =
{\bm 1}, 
\label{Eq_G_N}
\end{eqnarray}
where the Green function $\mathcal{ G }^{^S}$ is given by Eqs. (\ref{G^S_int}) and (\ref{g_f_S}), and $H^{^N}_{\bm k}$ is the bare surface Hamiltonian:
\begin{eqnarray}
H^{^N}_{\bm k}=
\left[
\begin{array}{cc}
h^{^N}_{\bm k} & 0    \\
0 & - h^{^{N*} }_{-{\bm k} }  
\end{array}
\right]. 
\label{H_N_0}
\end{eqnarray}

\subsection{ Clean TI}

Let us consider first Eq. (\ref{Eq_G_N}) in the absence of disorder: 
\begin{eqnarray}
[ \epsilon - H^{^N}_{\bm k} - \mathcal{T}^2 \tau_z\sigma_0 \mathcal{ G }^{^S}  \tau_z\sigma_0 ] G^{^{N0} }_{\bm k} = {\bm 1},
\label{Eq_G_N0}
\end{eqnarray}
where $G^{^{N0} }_{\bm k}$ denotes the Green function of a clean TI. 
Using Eqs. (\ref{G^S_int}) and (\ref{g_f_S}) for the momentum-integrated Green function $\mathcal{ G }^{^S}$, 
we can write Eq. (\ref{Eq_G_N0}) as follows 

\begin{eqnarray}
\left[
\begin{array}{cc}
\epsilon + i\Gamma_{_N}(\epsilon) - h^{^N}_{\bm k}    &  -\Delta_{_N}(\epsilon) i\sigma_y {\rm e}^{i\chi}  \\
\Delta_{_N}(\epsilon) i\sigma_y {\rm e}^{-i\chi}   &  \epsilon + i\Gamma_{_N}(\epsilon) + h^{^{N*} }_{ -{\bm k} }     
\end{array}
\right]
G^{^{N0} }_{\bm k} = {\bm 1},
\label{Eq_G_N0_1}
\end{eqnarray}
where the induced pairing potential $\Delta_{_N}(\epsilon)$ and the spectrum shift $i\Gamma_{_N}(\epsilon)$ are 
related to the momentum-integrated Green functions of the S:  
\begin{eqnarray}
&&
\Delta_{_N}(\epsilon) = i \Gamma_0 f_{_S}( \epsilon ) =  i \Gamma_0 \frac{\Delta_{_S}}{\sqrt{ \epsilon^2 - \Delta^2_{_S} }}, 
\label{Delta_N}\\
&&
\Gamma_{_N}(\epsilon) = \Gamma_0 g_{_S}( \epsilon ) = \Gamma_0 \frac{\epsilon}{\sqrt{ \epsilon^2 - \Delta^2_{_S} }},
\label{Gamma_N}\\
&&
\Gamma_0 =\pi \mathcal{T}^2 \nu_{_S}. 
\label{Gamma_0}
\end{eqnarray}
Here $\Gamma_0$ is the tunneling energy scale. It is the same scale that determines the normal-state level broadening due to  
the quasiparticle escape into S with the normal-state DOS $\nu_{_S}$.  
For large Fermi energy $E_{_N} \gg \Delta_{_N}$ and close to the TI Fermi momentum $k_{_N}$ 
the solution of Eq. (\ref{Eq_G_N0_1}) is given by
\begin{widetext}
\begin{eqnarray}
G^{ ^{N0} }_{\bm k}= \left[
\begin{array}{cc}
G^{ ^{N0} }_{ 11,{\bm k} }  &  G^{ ^{N0} }_{ 12,{\bm k} }  \\
G^{ ^{N0} }_{ 21,{\bm k} }  &  G^{ ^{N0} }_{ 22,{\bm k} }
\end{array}
\right]= 
\frac{
\frac{1}{2} 
\left[
\begin{array}{cc}
(\sigma_0 + \mbox{\boldmath$\sigma$}\cdot \hat{\bm k} ) \, ( \epsilon_{_N}(\epsilon) + \eta_{_N}(k) ) & 
(\sigma_0 + \mbox{\boldmath$\sigma$} \cdot \hat{\bm k} ) i\sigma_y\, \Delta_{_N}(\epsilon)\, {\rm e}^{i\chi}  \\
-i\sigma_y (\sigma_0 + \mbox{\boldmath$\sigma$}\cdot \hat{\bm k} )  \, \Delta_{_N}(\epsilon)\, {\rm e}^{-i\chi} & 
-i\sigma_y (\sigma_0 + \mbox{\boldmath$\sigma$} \cdot \hat{\bm k} )i\sigma_y \, ( \epsilon_{_N}(\epsilon) - \eta_{_N}(k) )
\end{array}
\right] 
}
{ 
\epsilon^2_{_N}(\epsilon) - \eta^2_{_N}(k) - \Delta^2_{_N}(\epsilon)  
},
\label{G^N0}
\end{eqnarray}
\end{widetext}
where the indices $1,2$ refer to the Nambu space, and functions $\epsilon_{_N}(\epsilon)$ and $\eta_{_N}(k)$ are defined by 
\begin{equation} 
\epsilon_{_N}(\epsilon) = \epsilon + i\Gamma_{_N}(\epsilon), \qquad \eta_{_N}(k) =\hbar v_{_N}(k-k_{_N}),
\label{e_N}
\end{equation}
with $v_{_N} = A_{_N}/\hbar$ denoting the Fermi velocity in the TI.
It is also convenient to introduce an alternative form of the Green function, $\tilde{G}^{ ^{N0} }_{\bm k}$, related to (\ref{G^N0}) by a unitary transformation $U$: 
\begin{equation}
G^{ ^{N0} }_{\bm k} = U \tilde{G}^{ ^{N0} }_{\bm k} U^\dagger, \quad 
U = \left[
\begin{array}{cc}
\sigma_0  &  0  \\
0 &  -i\sigma_y 
\end{array}
\right], 
\label{U}
\end{equation}
\begin{equation}
\tilde{G}^{ ^{N0} }_{\bm k} = \frac{1}{2}(\sigma_0 + \mbox{\boldmath$\sigma$}\cdot \hat{\bm k} ) \otimes C^{ ^0 }_{\bm k},
\label{G^N0_t}
\end{equation}
where the spin and particle-hole sectors are decomposed by the direct product $\otimes$, 
and $C^{ ^0 }_{\bm k}$ is the Green function in the Nambu space only
\begin{equation}
C^{ ^0 }_{\bm k} = 
\frac{
\left[
\begin{array}{cc}
\epsilon_{_N}(\epsilon) + \eta_{_N}(k)  & 
\Delta_{_N}(\epsilon)\, {\rm e}^{i\chi}  \\
\Delta_{_N}(\epsilon)\, {\rm e}^{-i\chi} & 
\epsilon_{_N}(\epsilon) - \eta_{_N}(k) 
\end{array}
\right] 
}
{ 
\epsilon^2_{_N}(\epsilon) - \eta^2_{_N}(k) - \Delta^2_{_N}(\epsilon).  
}
\label{C^0}
\end{equation}

\subsubsection{ Induced gap } 

Let us discuss Eq. (\ref{G^N0}). 
The single-particle excitations in the superconducting TI are described by the Green functions $G^{^{N0}}_{ 11,{\bm k} }$ and $G^{^{N0}}_{ 22,{\bm k} }$. 
Both of them involve the projector  
$\frac{1}{2}(\sigma_0 + \mbox{\boldmath$\sigma$}\cdot \hat{\bm k} )$,
indicating that the quasiparticles are $+1$ eigenstates of the helicity $\mbox{\boldmath$\sigma$}\cdot \hat{\bm k}$, 
just like in the normal state (cf. Ref. \onlinecite{GT11}). 
The pole of Eq. (\ref{G^N0}) yields the equation for the quasiparticle spectrum:
\begin{equation}
\epsilon^2_{_N}(\epsilon)  - \eta^2_{_N}(k) - \Delta^2_{_N}(\epsilon)  = 0.
 \label{Eq_Spectrum}
\end{equation}
For $k$ close to $k_{_N}$ the spectrum is
\begin{equation}
\epsilon_{\bm k} \approx \pm \sqrt{ \eta^2_{_N}(k) + \varepsilon^2_{\rm g} }.
\label{Spectrum}
\end{equation}
It has an induced energy gap $\varepsilon_{\rm g}$ which satisfies the equation $\epsilon_{_N}^2(\varepsilon_{\rm g})=\Delta_{_N}^2(\varepsilon_{\rm g})$. 
The latter can be explicitly written as (cf. Ref. \onlinecite{Kopnin11})  
\begin{equation}
\left( \frac{\varepsilon_{\rm g}}{\Delta_{_S}} \right)^2 \left[ 1 - \left( \frac{\varepsilon_{\rm g}}{\Delta_{_S}} \right)^2 \right] =
\gamma^2 \, 
\left( 
1-
\frac{\varepsilon_{\rm g}}{\Delta_{_S}}
\right)^2, \quad \gamma = \frac{\Gamma_0}{\Delta_{_S}},
\label{Eq_Eg}
\end{equation}
where we introduce a dimensionless parameter $\gamma$. 
Assuming $\gamma \ll 1$ for the purpose of this calculation,
we search for the solution to Eq. (\ref{Eq_Eg}) in the form of the expansion:
\begin{equation}
\frac{\varepsilon_{\rm g}}{\Delta_{_S}} = \gamma c_1 + \gamma^2 c_2 + \gamma^3 c_3 +... 
\label{Exp_gap}
\end{equation}
The constants $c_1, c_2, c_3 ...$ are obtained from comparing the coefficients at $\gamma, \gamma^2, \gamma^3 ...$ 
on the left- and right-hand sides of Eq. (\ref{Eq_Eg}). 
Up to the cubic terms we find $c_1 = -c_2 = 1$ and $c_3 = 3/2$, which yields the induced gap 
\begin{eqnarray}
\varepsilon_{\rm g} \approx \Delta_{_S} \left(\gamma - \gamma^2 + \frac{3}{2} \gamma^3\right) = \Gamma_0 \left(1 - \gamma + \frac{3}{2} \gamma^2\right). 
\label{Eg}
\end{eqnarray}
This equation shows that $\varepsilon_{\rm g}$ is reduced as the gap in the superconductor, $\Delta_{_S}$, becomes smaller.
It is also worth noting that the energy $\Gamma_0$ can be expressed in terms of experimentally accessible parameters of a S/TI interface: \cite{Maier12}
\begin{equation}
 \Gamma_0 = \frac{h g_n}{2 e^2} \, \frac{ \hbar v }{ k_F }, \qquad k_F = \sqrt{ 4\pi n },
\label{Gamma0}
\end{equation}
where $g_n$ is the normal-state interface conductance per unit area and $k_F$ is the Fermi wave-number of the TI surface state 
determined by the surface carrier density. The energy scale $\Gamma_0$ can be extracted 
from the temperature dependence of the critical current in short proximity-effect junctions.~\cite{Rohlfing09} 

\subsubsection{ Helical $s + p$ - wave pair correlations } 

The off-diagonal entry $G^{^{N0}}_{ 21,{\bm k} }$ (as well as $G^{^{N0}}_{ 12,{\bm k} }$) in Eq. (\ref{G^N0})
is the Green function of the induced superconducting condensate. 
Its structure in spin space reveals the symmetry of the induced pair correlations:
\begin{equation}
G^{^{N0}}_{ 21,{\bm k} } \propto i\sigma_y (\sigma_0 + {\bm\sigma}\cdot \hat{\bm k} ) \, \Delta_{_N}(\epsilon).
\label{Pair}
\end{equation}
It is a mixture of the singlet $s$-wave component (first term) and a triplet $p$-wave component (second term, see e.g. Ref. \onlinecite{Leggett75}).
Equation (\ref{Pair}) agrees with the results of Ref. \onlinecite{Stanescu10} for a large Fermi energy. 
The origin of the mixed $s$- and $p$-wave superconducting correlations is the broken spin-rotation symmetry 
of the helical surface states. In this sense the situation is similar to the mixed singlet-triplet intrinsic superconductivity predicted 
for systems without inversion symmetry. \cite{Gorkov01,Santos10} 
We note that the induced $p$-wave component inherits the spin-momentum locking $\mbox{\boldmath$\sigma$} \cdot \hat{\bm k}$ 
of the normal-state carriers.

\subsection{ Disordered TI }

In the presence of disorder it is convenient to recast Eq. (\ref{Eq_G_N}) in the Dyson form:
\begin{equation}
G^{ ^{N} }_{\bm k} = G^{ ^{N0} }_{\bm k} + u^2 \tau_z\sigma_0 \mathcal{ G }^{^N}  \tau_z\sigma_0 \, G^{ ^{N} }_{\bm k},
\label{Eq_G_N_Dyson}
\end{equation}
where $G^{ ^{N0} }_{\bm k}$ includes the tunneling and is given by Eq. (\ref{G^N0}). 
Guided by the decomposition (\ref{U}) and (\ref{G^N0_t}) we seek the solution 
to Eq. (\ref{Eq_G_N_Dyson}) in the form 
\begin{equation}
G^{ ^N }_{\bm k} = U \tilde{G}^{ ^N }_{\bm k} U^\dagger, \,\, 
\tilde{G}^{ ^N }_{\bm k} = \frac{1}{2}(\sigma_0 + \mbox{\boldmath$\sigma$}\cdot \hat{\bm k} ) \otimes C_{\bm k},
\label{U1}
\end{equation}
where $C_{\bm k}$ is a $2\times 2$ Green function in the Nambu space, for which we derive the following equation:
\begin{equation}
 C_{\bm k} = C^{^0}_{\bm k} + u^2 C^{^0}_{\bm k} \int \frac{d{\bm q}}{(2\pi)^2} \frac{1 + \hat{\bm k} \cdot \hat{\bm q}  }{2} \tau_z C_{\bm q} \tau_z  C_{\bm k},
\label{Eq_C}
\end{equation}
with $C^{^0}_{\bm k}$ is given by Eq. (\ref{C^0}). The particle helicity results in the anisotropic kernel $(1 + \hat{\bm k} \cdot \hat{\bm q} )/2$ 
depending on the scattering angle between the directions of the initial $\hat{\bm k}$ and final $\hat{\bm q}$ momentum states. 
Since Eq. (\ref{Eq_C}) contains no spin degrees of freedom, it can be solved by analogy with the case of the conventional S in Sec. \ref{S}, 
which yields
\begin{equation}
C_{\bm k} = 
\frac{
\left[
\begin{array}{cc}
\epsilon^\prime_{_N} + \eta_{_N}(k)  & 
\Delta^\prime_{_N} \, {\rm e}^{i\chi}  \\
\Delta^\prime_{_N} \, {\rm e}^{-i\chi} & 
\epsilon^\prime_{_N} - \eta_{_N}(k) 
\end{array}
\right] 
}
{ 
\epsilon^{\prime 2}_{_N} - \eta^2_{_N}(k) - \Delta^{\prime 2}_{_N}.  
}
\label{C}
\end{equation}
Here $\epsilon^\prime_{_N}(\epsilon, \hat{\bm k})$ and $\Delta^\prime_{_N}(\epsilon, \hat{\bm k})$ 
are functions of the energy $\epsilon$ and the momentum direction $\hat{\bm k}$ 
satisfying the self-consistency equations [cf. Eqs. (\ref{Delta_S}) and (\ref{E_S})]:
\begin{eqnarray}
&&
\epsilon^\prime_{_N} =\epsilon_{_N}(\epsilon) + \frac{i\hbar}{ 2\tau_{_N} } 
\overline{
\frac{1 + \hat{\bm k}\cdot \hat{\bm q}}{2} 
\frac{  \epsilon^\prime_{_N}  }{  \sqrt{ \epsilon^{\prime 2}_{_N} - \Delta^{\prime 2}_{_N}  }   }
},
\label{E_N}\\
&&
\Delta^\prime_{_N} = \Delta_{_N}(\epsilon) + \frac{i\hbar}{ 2\tau_{_N} } 
\overline{
\frac{1 + \hat{\bm k}\cdot \hat{\bm q} }{2} 
\frac{  \Delta^\prime_{_N}  }{  \sqrt{ \epsilon^{\prime 2}_{_N} - \Delta^{\prime 2}_{_N}  }  }
}. 
\label{D_N}
\end{eqnarray}
The bar denotes averaging $\overline{( ...)} = \int^{2\pi}_0\frac{  d\phi_{ \hat{\bm q} }  }{2\pi} (...) $  
over the momentum direction $\hat{\bm q}$ of the final state on the Fermi surface and the time-scale 
\begin{equation}
\tau_{_N}=\hbar/(2 \pi u^2 \nu_{_N} )
\label{tau_N}
\end{equation}
is determined by the disorder strength $u$ and the normal-state DOS per spin in the TI, $\nu_{_N}$. 

Returning to the full Green function (\ref{U1}) we notice that it retains the structure of the disorder-free Green function (\ref{G^N0}), 
with $\Delta_{_N}$ and $\epsilon_{_N}$ replaced by $\Delta^\prime_{_N}$ (\ref{D_N}) and $\epsilon^\prime_{_N}$ (\ref{E_N}):   
\begin{widetext}
\begin{eqnarray}
G^{ ^{N} }_{\bm k} = \left[
\begin{array}{cc}
G^{ ^{N} }_{ 11,{\bm k} }  &  G^{ ^{N} }_{ 12,{\bm k} }  \\
G^{ ^{N} }_{ 21,{\bm k} }  &  G^{ ^{N} }_{ 22,{\bm k} }
\end{array}
\right]= 
\frac{
\frac{1}{2} 
\left[
\begin{array}{cc}
(\sigma_0 + \mbox{\boldmath$\sigma$}\cdot \hat{\bm k} ) \, ( \epsilon^\prime_{_N} + \eta_{_N}(k) ) & 
(\sigma_0 + \mbox{\boldmath$\sigma$}\cdot \hat{\bm k} ) i\sigma_y \, \Delta^\prime_{_N}\, {\rm e}^{i\chi}  \\
-i\sigma_y (\sigma_0 + \mbox{\boldmath$\sigma$}\cdot \hat{\bm k} )  \, \Delta^\prime_{_N}\, {\rm e}^{-i\chi} & 
-i\sigma_y (\sigma_0 + \mbox{\boldmath$\sigma$}\cdot \hat{\bm k} ) i\sigma_y\, ( \epsilon^\prime_{_N} - \eta_{_N}(k) )
\end{array}
\right] 
}
{ 
\epsilon^{\prime 2}_{_N} - \eta^2_{_N}(k) - \Delta^{\prime 2}_{_N}  
}. 
\label{G^N}
\end{eqnarray}
\end{widetext}

\subsubsection{Solution of the self-consistency equations}

We seek isotropic solutions $\epsilon^\prime_{_N}(\epsilon, \hat{\bm k})=\epsilon^\prime_{_N}(\epsilon)$ and 
$\Delta^\prime_{_N}(\epsilon, \hat{\bm k})=\Delta^\prime_{_N}(\epsilon)$ for which Eqs. (\ref{E_N}) and (\ref{D_N}) can be written as 
\begin{eqnarray}
&&
\epsilon^\prime_{_N} - \frac{i\hbar}{ 2\tau } \frac{ \epsilon^\prime_{_N} }{ \sqrt{ \epsilon^{\prime 2}_{_N} - \Delta^{\prime 2}_{_N}  } } 
= \epsilon_{_N},
\label{E_N1}\\
&&
\Delta^\prime_{_N} - \frac{i\hbar}{ 2\tau } \frac{ \Delta^\prime_{_N} }{ \sqrt{ \epsilon^{\prime 2}_{_N} - \Delta^{\prime 2}_{_N}  } } =  
\Delta_{_N},
\quad 
\tau = 2\tau_{_N}. 
\label{D_N1}
\end{eqnarray}
These equations have the same form as in the isotropic S [cf. Eqs. (\ref{Delta_S}) and (\ref{E_S})] except that 
the elastic life-time $\tau$ acquires an extra factor of $2$ due to the scattering anisotropy. 
Dividing Eq. (\ref{E_N1}) by Eq. (\ref{D_N1}) we find 
\begin{equation}
\frac{ \epsilon^\prime_{_N} }{ \Delta^\prime_{_N} } = \frac{ \epsilon_{_N} }{ \Delta_{_N} },
\label{E_to_D}
\end{equation}
which upon substituting back to Eqs. (\ref{E_N1}) and (\ref{D_N1}) yields the explicit solutions:
\begin{eqnarray}
&&
\epsilon^\prime_{_N} = \epsilon_{_N} \left(  1 + \frac{i\hbar}{ 2\tau } \frac{1}{ \sqrt{ \epsilon^2_{_N} - \Delta^2_{_N}  } } \right),
\label{E_N2}\\
&&
\Delta^\prime_{_N} =  \Delta_{_N} \left(  1 + \frac{i\hbar}{ 2\tau } \frac{1}{ \sqrt{ \epsilon^2_{_N} - \Delta^2_{_N}  } } \right). 
\label{D_N2}
\end{eqnarray} 
These equations along with Eq. (\ref{G^N}) provide the self-consistent solution for the Green function of a disordered TI.

\subsubsection{ DOS }

\begin{figure}[t]
\begin{center}
\includegraphics[width=65mm]{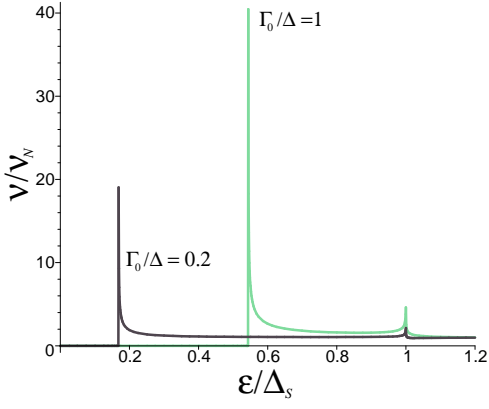}
\end{center}
\caption{
Normalized density of states in TI [see Eq. (\ref{DOS_2})] for different interface parameters.
}
\label{DOS_fig}
\end{figure}

As an application of the self-consistent Green function solution, we now briefly discuss the surface DOS $\nu(\epsilon)$. It can be calculated from the formula:   
\begin{equation}
 \nu(\epsilon) = -\frac{1}{\pi} \int \frac{ d{\bm k} }{ (2\pi)^2 } {\rm Im \, Tr} \, G^{ ^{N \, R} }_{11,\bm k}( \epsilon ),
\label{DOS_def} 
\end{equation}
where $G^{ ^{N\, R} }_{11,\bm k}( \epsilon )$ is the retarded particle Green function which coincides with the $G^{ ^{N} }_{11,\bm k}( \epsilon )$ block of Eq. (\ref{G^N}) 
taken for $\epsilon > 0$, ${\rm Im }$ denotes the imaginary part and ${\rm Tr }$ is the trace in spin space. After integration over the momentum, we have 
\begin{eqnarray}
\nu(\epsilon) &=& \nu_{_N}{\rm Re } \,\frac{ \epsilon^\prime_{_N}(\epsilon) }{ \sqrt{  \epsilon^{\prime 2}_{_N}(\epsilon) - \Delta^{\prime 2}_{_N}(\epsilon)  }  } =
\label{DOS_1}\\
&=&
\left\{
\begin{array}{cc}
0,  &  \epsilon < \varepsilon_{\rm g},  \\
\nu_{_N} \frac{ \epsilon_{_N}(\epsilon) }{ \sqrt{  \epsilon^2_{_N}(\epsilon) - \Delta^2_{_N}(\epsilon)  }  },  &  \epsilon > \varepsilon_{\rm g},
 \end{array}
\right.
\label{DOS_2} 
\end{eqnarray}
where ${\rm Re}$ denotes the real part. We notice that the disorder-dependent factors in the numerator and denominator of Eq. (\ref{DOS_1}) cancel out, 
and the DOS assumes the usual form for the conventional proximity systems (see e.g., Refs. \onlinecite{Fagas05} and \onlinecite{Aminov96}, and in Fig. \ref{DOS_fig}).
It features two peaks at energies corresponding to the induced gap $\varepsilon_{\rm g} < \Delta_{_S}$ and to the gap of the superconducting material $\Delta_{_S}$.   
The robustness of the DOS with respect to disorder is due to the fact that only the isotropic part of the Green function (\ref{G^N}) contributes to Eq. (\ref{DOS_def}).   

\subsubsection{Condensate correlation function}

At energies below the induced gap $\varepsilon_{\rm g}$, when no single-particle excitations exist, 
the system is a superconducting condensate that exhibits pair correlations characterized by the anomalous Green function $G^{ ^{N} }_{21,\bm k}( \epsilon )$ 
in Eq. (\ref{G^N}). In particular, the anomalous Green function contains information on how fast the superconducting correlations decay with distance 
${\bm R}$ in real space. The real-space correlation function can be defined as the Fourier transform:
\begin{equation}
G^{ ^{N} }_{21}(\epsilon, {\bm R} ) = \int \frac{ d{\bm k} }{(2\pi)^2} \, G^{ ^{N} }_{21,\bm k}( \epsilon ) \, {\rm e}^{i{\bm k}{\bm R}}.
\label{Corr_def}
\end{equation}
The momentum integration can be done in two steps. First, the integration over the absolute value $|{\bm k}|=k$ is performed. 
To do so we change the integration variable to $\eta_{_N}(k)$ and extend the lower integration limit to $-\infty$, 
which is permissible for large Fermi energies $E_{_N} \gg \Delta^\prime_{_N}$. 
Then, we do the integration over the momentum direction $\hat{\bm k}$ which is specified by angle $\phi$ in the following equations.   
The resulting expression for $G^{ ^{N} }_{21}(\epsilon, {\bm R} )$ is
\begin{eqnarray}
&&
G^{ ^{N} }_{21}(\epsilon, {\bm R} ) = \frac{ \pi \nu_{_N} \Delta_{_N}(\epsilon) {\rm e}^{-i\chi} }{ 2\sqrt{  \Delta^2_{_N}(\epsilon) - \epsilon^2_{_N}(\epsilon)  }  }
\label{Corr}\\
&&\times
\left[ 
i\sigma_y F^{(s)}(R) -
\sigma_y \mbox{\boldmath$\sigma$}\cdot \frac{ {\bm R} }{R} \, F^{ (p) }(R) 
\right], \quad R = |{\bm R}|,
\nonumber
\end{eqnarray}
where $F^{(s)}(R)$ and $F^{ (p) }(R)$ are the dimensionless amplitudes of the $s$- and $p$-wave pair correlations given by
\begin{eqnarray}
F^{(s)}(R) &=& \frac{2}{\pi} \int\limits_0^{\pi/2} \cos( k_{_N} R \cos\phi ) \times
\nonumber\\
&\times& 
\exp\left( - \left[ \frac{1}{\xi} + \frac{1}{2\ell}   \right] R \cos\phi \right) d\phi, 
\label{F^s}
\end{eqnarray}
\begin{eqnarray}
F^{ (p)}(R) &=& \frac{2}{\pi} \int\limits_0^{\pi/2} \cos\phi \sin( k_{_N} R \cos\phi ) \times 
\nonumber\\
&\times& 
\exp\left( - \left[ \frac{1}{\xi} + \frac{1}{2\ell}   \right] R \cos\phi \right) d\phi. 
\label{F^p}
\end{eqnarray}
These functions depend on the absolute value of the distance $R$, the Fermi momentum of the TI $k_{_N}$, the coherence length of a clean TI, $\xi$, 
and the elastic mean-free path, $\ell$. The latter two are defined by 
\begin{eqnarray}
\xi = \frac{\hbar v_{_N}}{ \sqrt{  \Delta^2_{_N}(\epsilon) - \epsilon^2_{_N}(\epsilon)  }  }, 
\quad
\ell = v_{_N} \tau =2v_{_N}\tau_{_N}.
\label{xi_ell}
\end{eqnarray}

In accord with the Pauli exclusion principle the spin-triplet $p$-wave correlations (\ref{F^p}) are nonlocal, with the amplitude $F^{(p)}(R)$ vanishing at $R=0$. 
Due to their inherent nonlocality the $p$-wave correlations depend stronger on the elastic mean-free path $\ell$, 
which leads to the suppression of the $p$-wave component in disordered samples. This is demonstrated in Fig. \ref{F_p_fig} for a fixed value of parameter $k_{_N}\xi = 10$, 
which correponds to the large-Fermi-energy regime. 
The disorder strength is characterized by the ratio $\xi/\ell$ varying from $0$ in a clean system with $\ell \to \infty$ (curve a in Fig. \ref{F_p_fig}) 
to $\xi/\ell \gg 1$ in a dirty sample with a short mean-free path (curve c in Fig. \ref{F_p_fig}). 
The comparison of curves a and c in Fig. \ref{F_p_fig} shows that the disorder-induced suppression of the $p$-wave correlations is indeed significant.
In contrast, the $s$-wave amplitude (\ref{F^s}) is finite for zero distance, $F^{(s)}(0)=1$ independently of disorder (see also Fig. \ref{F_s_fig}). 
The effect of a finite mean-free path becomes visible only with inceasing $R$. Still, it is much weaker than in the $p$-wave case. 
This is clearly seen from the asymptotics of Eqs.(\ref{F^s}) and (\ref{F^p}) at large distances $R \gg 2\ell\xi/( 2\ell + \xi )$:
\begin{equation}
F^{(s)}(R) \approx \frac{2}{\pi R} \frac{2\ell\xi}{ 2\ell + \xi  }= 
\left\{
\begin{array}{cc}
\frac{2}{\pi} \frac{\xi}{R},  &  \xi \ll \ell,  \\
 &  \\
\frac{4}{\pi} \frac{ \ell }{R}, &  \xi \gg \ell,
 \end{array}
\right.
\label{F^s_1}
\end{equation}
\begin{equation}
F^{(p)}(R) \approx \frac{ 4k_{_N} }{\pi R^2} \frac{ (2\ell\xi)^3 }{ ( 2\ell + \xi )^3 }= 
\left\{
\begin{array}{cc}
\frac{4}{\pi} \frac{ k_{_N} \xi^3}{R^2},  &  \xi \ll \ell,  \\
 &  \\
\frac{32}{\pi} \frac{ k_{_N} \ell^3 }{R^2}, &  \xi \gg \ell.
 \end{array}
\right.
\label{F^p_1}
\end{equation}
Unlike the $s$-wave amplitude (\ref{F^s_1}), the $p$-wave one (\ref{F^p_1}) 
is proportional to the third power of the mean-free path $\ell$ in the dirty case 
$\xi \gg \ell$.

Thus, the main conclusion of this study is that in dirty TIs, in which the elastic mean-free path 
is much smaller than the superconducting coherence length ($\ell \ll \xi$),  
the induced superconductivity is predominantly $s$-wave like. The $p$-wave component is greatly suppressed compared with the $s$-wave one.  
In cleaner TIs the $p$-wave amplitude remains always smaller than the $s$-wave one, 
albeit the difference between them may not be so drastic (see curves b for $\xi/\ell =5$ in Figs. \ref{F_p_fig} and \ref{F_s_fig}).
Both components should be observable in the presence of a modest amount of impurities.  
It is worth mentioning that in this work the tunneling between a superconductor and a TI has been treated as an incoherent process 
in which the electron momentum is not conserved. This case applies to spatially nonuniform interfaces. 
However, similar results for the Green functions, the DOS and condensate correlation functions are expected for coherent tunneling. 
The self-consistent Green functions derived in this paper can be used to study superconducting transport in impure TIs.  

\acknowledgments
This work was supported by the German research foundation (DFG), Grants No FOR1162 (HA5893/5-2) and TK60/1-1.


\begin{thebibliography}{99}
\bibitem{Fermi}
In equilibrium the electrochemical potentials in the S and TI are equal, but 
in view of the different band structures and, hence, different electron fillings of the conduction bands in the S and TI
the Fermi energies $E_{_S}$ and $E_{_N}$ need not be identical.
\end{thebibliography}
\end{document}